\documentstyle[onecolumn]{mn}
\def\beq{\begin{equation}}
\def\eeq{\end{equation}}
\def\bey{\begin{eqnarray}}
\def\eey{\end{eqnarray}}

\input epsf

\begin{document}
\title
[Gravitational lensing models of PG1115+080]
{Systematic uncertainties in gravitational lensing models: a semi-analytical study of PG1115+080}
\author[H. Zhao \& D. Pronk]
{HongSheng Zhao \& Danny Pronk
\thanks{E-mail: hsz@strw.leidenuniv.nl, pronk@strw.leidenuniv.nl}  
                       \\ 
Sterrewacht Leiden, Niels Bohrweg 2, 2333 CA, Leiden, The Netherlands\\
}
\date{Accepted ........      Received .......;      in original form .......}
\pagerange{\pageref{firstpage}--\pageref{lastpage}}\pubyear{2000}\maketitle
\label{firstpage}

\begin{abstract}
While the Hubble constant can be derived from observable time delays
between images of lensed quasars, the result is often highly sensitive
to assumptions and systematic uncertainties in the lensing model.  
Unlike most previous authors we put minimal restrictions on
the radial profile of the lens and allow for non-elliptical lens potentials.
We explore these effects using a broad class of models with a lens potential 
$\psi(r f(\theta))$, which has an unrestricted radial profile
but self-similar iso-potential contours defined by $r f(\theta) =$ constant.  
For these potentials the lens equations can be solved semi-analytically.
The axis ratio and position angle of the lens
can be determined from the image positions of quadruple
gravitational lensed systems directly, independent of the radial
profile.  We give simple equations for estimating the power-law
slope of the lens density directly from the image positions and for
estimating the time delay ratios.  Our method greatly simplifies the
numerics for fitting observations and is fast in exploring the model
parameter space. As an illustration we apply the model
to PG1115+080. An entire one-parameter sequence of models fit the
observations exactly. We show that the measured image positions and
time delays do not uniquely determine the Hubble constant.
\end{abstract}


\section{Introduction}

Gravitationally lensed systems are powerful probes of galactic
potentials and the scale of the universe.  The advantage over the
traditional stellar dynamical method is that from the image positions
we can measure the shape and the mass of the dark halo well beyond the
half-light-radius of a faraway lens galaxy, whether it is virialized
or not.  The time delay between two images is a measurement of the
difference in the length of the two bent light paths, and scales
with the distances to the lens and source.  So the Hubble constant can
be constrained once the redshift of the lens and the source and the
time delay are measured.  This way of getting $H_0$ has the advantage
that the underlying physics (general relativity) can be rigorously
modeled.  The limitation is that there is often a sequence of lens
models that can fit the image positions.

Presently about twenty strongly lensed systems are known, half of them being
quadruple-imaged quasars and half being double-imaged quasars (e.g.,
Keeton \& Kochanek 1996).  We will concentrate on quadruple-imaged
systems.  They are better constrained than double-imaged systems,
since the lens model needs to fit more image positions and also the ratios
of time delays between any two pairs of images.  Quadruple systems
typically involve a quasar source well-aligned with the center of the
lens potential well.  Presently only two such systems (PG1115+080 and 
B1608+656) have accurately determined image positions and time delays.

A significant amount of numerical computation is usually required to
invert image positions to intrinsic parameters of the potential
(cf. Schneider, Ehlers \& Falco 1992).  The degeneracy of the
resulting potential is often not fully explored because of the need to
cover a large parameter space, particularly for flattened potentials.
Previous authors have often restricted their studies to isothermal or
power-law spherical models (Evans \& Wilkinson 1998 and references
therein), and elliptical models (Witt \& Mao 1997 and references
therein) and other simple models (Kassiola \& Kovner 1993, 1995) with
or without external shear.  The fully general non-parametric method,
e.g., the pixelated lens method of Williams \& Saha (2000), is very
powerful in demonstrating the complete range of the degeneracy in the
lens models, but it involves significant amount of numerical
computation and does not provide a clear insight to the relations
between the characteristic parameters of the lens.  For these reasons
it is still desirable to find analytical, yet general potentials,
which allow a quick exploration of the model parameter space.  For
example, it is of interest to generalize the analytical work of Witt
\& Mao (1997) to non-elliptical lenses.  It would also be interesting
to find analytical expressions for the time delay in these general lenses,
which could help us to understand how
the radial profile and lens shape affect the predictions on the Hubble
constant.

Here we study a broad class of analytical models with
non-axisymmetric, non-elliptical 
shape and semi-power-law radial profile (\S 2), and show how to
calculate the lens shape and radial profile parameters directly from
the image positions (\S 3).  We apply the models to PG1115+080 (\S 4)
and show that the images can be fit perfectly by a large range of lens
models (\S 5).  We summarize our results in \S 6 and conclude with the
implications on the Hubble constant.

\section{Lens equation in a general class of models}

\subsection{Decoupling of angular dependence and radial profile}

Any two-dimensional lens potential can be cast in the following
form,
\beq
\psi(x,y)=\psi(\omega),~~~\omega=\omega(r,\theta),
\eeq
where $\psi$ has the dimension of square arcsec, $(x,y)$ defines a
rectangular coordinate system (in units of arcsecs to the West and
North of the lens galaxy center) and $(r,\theta)$ the corresponding
polar coordinate system with 
\beq
(x, y)=(-r \sin\theta, r \cos\theta),
\eeq
where $\theta$ is the position angle, counterclockwise from North.
Unless otherwise specified we shall follow the notations of Schneider
et al. (1992).  
Here $\omega$ is defined to have the dimension of radius, so that 
constant $\omega$ curves correspond to equal-potential contours,
and define the shape and flattening of the potential.
The radial profile of the potential is a smooth function
$\psi(\omega)$ of the radius $\omega$.
We can also define
\beq\label{mdef}
m(\omega) \equiv \omega {d\psi \over d\omega}
\eeq
as the mass (in units of square arcsec) 
enclosed inside the radius $\omega$ (in units of arcsec).
For example, we have 
$m = \alpha \psi$ for a power-law model $\psi \propto \omega^\alpha$
with slope $\alpha$.
For a source at redshift $z_s$ and angular distance $D_{os}(z_s)$,
and a lens at redshift $z_l$ and distance $D_{ol}(z_l)$, the physical
mass $M(\omega)$ is related to $m(\omega)$ by the scaling
\beq
M(\omega) \equiv \pi \Sigma_c ~ m(\omega),
\eeq
where
\beq
\Sigma_c \equiv \frac{c^2}{4\pi G} \frac{D_{ol} D_{os}}{D_{ls}}
 \left(\frac{1 ~\rm radian}{206265 ~\rm arcsec}\right)^2 \approx
 39  \frac{D_{ol} D_{os}}{D_{ls}} 
\eeq
is the critical density in units of $M_\odot / \rm arcsec^2$, and
$D_{ls}(z_l,z_s)$ is the relative distance of the lens and source
and all distances are in units of parsec.  

A light ray from a source at $(r_s,\theta_s)$, being deflected to a direction
$(r,\theta)$ by the lensing galaxy with potential $\psi(\omega)$
located at the origin, will experience a time delay $\Delta t$ given by
\beq\label{tdelay}
\Delta t(r,\theta) =h^{-1} \tau_{100}(z_l, z_s) \left[{1 \over 2}r^2-r r_s \cos(\theta-\theta_s) + {1 \over 2}r_s^2 -\psi(\omega)\right],
\eeq
where $h$ is the Hubble constant $H_0$ rescaled to $100$~km/s/Mpc. A
characteristic value for the time delay is
\beq
\tau_{100}(z_l, z_s) \equiv {D_{ol} D_{os} \over D_{ls}}{1+z_l \over c},~~~\mbox{for $H_0=100$~km/s/Mpc}.
\eeq
According to Fermat's principle,
the images lie at the minimum of $\Delta t$, so the lens equation is given by
\bey\label{tr}
{\partial \over \partial r} \Delta t &=& r-r_s \cos(\theta-\theta_s) - m(\omega) \partial_r \ln\omega=0\\\label{ta}
{\partial \over r\partial \theta} \Delta t &=&r_s \sin(\theta-\theta_s) - {m(\omega) \over r} \partial_\theta \ln\omega=0.
\eey

Interestingly, the radial part $m(\omega)$ can be eliminated by
simply combining the two equations, 
\beq\label{comb}
{ r_s \sin(\theta-\theta_s) \over r - r_s \cos(\theta-\theta_s)}
={\partial_\theta \ln\omega \over r \partial_r \ln\omega},
\eeq
similar as in Witt \& Mao (1997).
This implies that there is a
relation linking the image positions to the shape of the potential
directly, {\it independent of the radial profile}.  

\subsection{Property of the image positions: the semi-hyperbolic curve}

For simplicity we shall concentrate on self-similar models with 
\beq
\label{omegadef}
\omega(r,\theta)=r \cdot f(\theta),
\eeq
where $f(\theta)$ defines the shape of the equal potential contours.
In principle the shape function can be
bi-symmetric or lopsided as long as the corresponding surface density
is positive everywhere in the lens plane.  
To be specific, we will restrict our discussions to 
bi-symmetric potentials with an angular part
\beq
\label{ftheta}
f(\theta)=\left| 1- \delta \cos 2\theta'\right|^\beta,
\eeq
which is a three-parameter 
$(\beta,\delta,\theta_p)$ function of the angle $\theta$,
where $\beta$ is a constant, the parameter $\delta$ is a flattening indicator,
and $\theta_p$ is the position angle of a principal axis of the potential.
The angle $\theta'$ is the azimuthal angle $\theta$ except for a rotation with
\beq\label{thetaprime}
\theta' \equiv \theta-\theta_p,~~~(x', y')=(r \cos\theta', r \sin\theta')
\eeq
so that $(x', y')$ defines a rectangular coordinate system with the axes 
coinciding with the principal axes of the lens.  For example, the 
source at
\beq
(x_s,y_s)=(-r_s\sin\theta_s,r_s\cos\theta_s)
\eeq
in the original rectangular coordinate system would be at 
\beq\label{xysp}
(x'_s, y'_s) =(r_s \cos\theta'_s, r_s \sin\theta'_s),
~~~\theta'_s \equiv \theta_s - \theta_p,
\eeq
in the rotated rectangular coordinate system.

With these we can compute the 
right-hand part of eq.~\ref{comb},
\beq\label{ut2}
{\partial_\theta \ln\omega \over r \partial_r \ln\omega}
={d \ln f(\theta) \over d\theta} = 
{1 \over v \cot \theta' + u\tan \theta'},
\eeq
where
\beq\label{uvdef}
u \equiv {1+\delta \over 2\Delta},~~~v \equiv {1- \delta \over 2\Delta},~~~\Delta=2\beta \delta,
\eeq
are shape indicators just like $\beta$ and $\delta$.
Substituting in eq.~\ref{comb}, and rewriting the image radius $r$ 
as a function of $\theta'$, we find the images fall on a family of curves 
$r=r(\theta')$ defined by 
\beq\label{curve}\label{rs}
r(\theta')= r_s \left[\cos(\theta'-\theta'_s)+ \sin(\theta'-\theta'_s) 
\left( v \cot \theta' + u\tan \theta' \right)
\right].
\eeq 
An alternative expression for these curves can be obtained by
expanding the sinusoidal terms in eq.~\ref{rs} so that
\beq\label{rxy}
r = (X_a \cos\theta'+Y_a \sin\theta')
   +\left({X_b \over \cos\theta'}
         -{Y_b \over \sin\theta'}\right),
\eeq
which is now linear in four new parameters $(X_a,Y_a,X_b,Y_b)$; 
these parameters
are related to the lens shape parameters $(v,u)$ and 
the source position $(x'_s,y'_s)$ by
\beq\label{XY}
X_a \equiv \left(1+v-u\right) x'_s,~~~
Y_a \equiv \left(1-u+v\right) y'_s,~~~
X_b \equiv u x'_s,~~~
Y_b \equiv v y'_s.
\eeq

These curves (cf. eq.~\ref{rs}) have the nice property that they go
through all image positions independent of the radial profile of the
lensing galaxy.  An example is the semi-hyperbolic curve in Fig.~\ref{surf2}.
The curve is determined by the source parameters $(r_s, \theta'_s)$,
the lens shape parameters $(v,u)$ and the lens 
position angle $\theta_p$.
The radial profile can take any general physical
profile, isothermal or power-law.

\begin{figure}
\epsfysize=10cm 
\centerline{\epsfbox{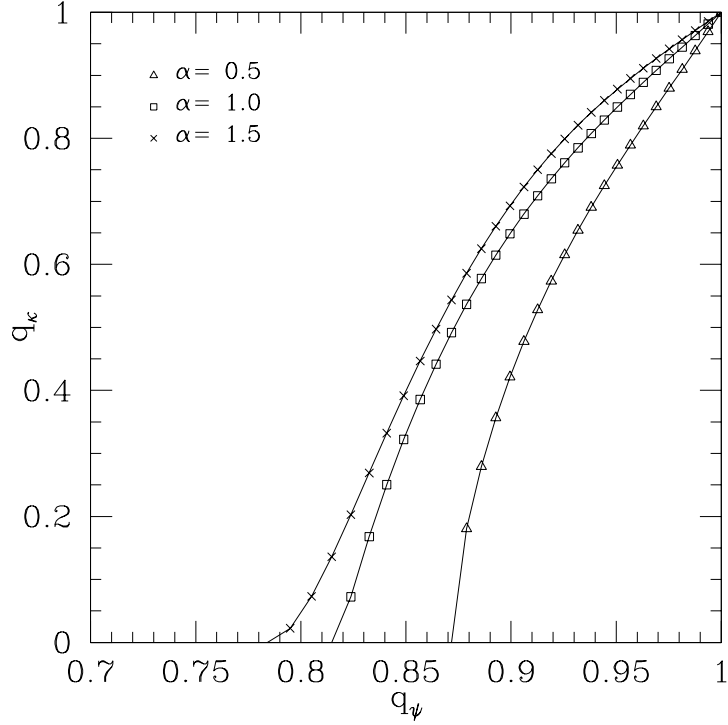}}
\caption{The axis ratio of the potential vs. that of the density
for our models with $\beta=-{1 \over 8}$ and 
the power-law index $\alpha=1.5$ (crosses),
1 (boxes) and 0.5 (triangles).
}\label{qq}
\end{figure}

The boxiness parameter $\beta$ is such that the shape function $f(\theta)$
reduces to the usual elliptical form when $\beta={1 \over 2}$ (Witt \&
Mao 1997).  In the case that $\psi(\omega)$ is linear in $\omega$, the
models reduce to the simple models of Kassiola \& Kovner (1995) when
$\beta=1$.  Interestingly, elliptical models with $\beta={1 \over 2}$
have $u-v={\delta \over \Delta}={1 \over 2\beta}=1$ (cf. eq. \ref{uvdef}),
hence
$X_a=Y_a=0$, and eq.~\ref{rxy} reduces to
\beq
1 = \frac{X_b}{x'} - \frac{Y_b}{y'}
\eeq
after applying $x'=r \cos\theta', y'=r \sin\theta'$. This equation
prescribes a hyperbolic curve, which  is
consistent with Witt (1996) and Witt \& Mao's (1997) finding that all
four image positions and the source position lie on a certain
hyperbolic curve.  A hyperbolic curve has a maximum of 5 free
parameters, thus they cannot fit image positions of a general
quadruple system; four image positions yield a minimal of eight
constraints.
Experience with fitting several quadruple systems (G2237+0305,
CLASS1608+656, HST12531-2914) shows that models with $|\beta| \ge 1/2$ often
give unphysical mass-radius relations $m(\omega)$.  We find that
$\beta=-{1 \over 8}$ gives a fair approximation to realistic models.  
These models have non-elliptical contours, and often yield physical density
distributions.  Fig.~\ref{qq} shows that they also cover a
sufficiently wide range of axis ratios for the potential and the density
so that we can explore the shape of the lens
galaxy in fitting the image positions.  The expressions for the axis
ratios of a power-law lens are given in Appendix A.

\section{Results}

\subsection{Lens shape directly from fitting image positions}

Our models can be used to fit image positions and derive lens shape
parameters ($\delta,\beta$) and the source position $(r_s,\theta_s)$,
free from assumptions of the lens radial profile, but subject to the
assumption that the lens' angular profile obeyes eq.~\ref{ftheta}.
The four unknowns can be derived from the four observed image
positions, i.e., the \emph{eight} observables $(r_i,\theta_i)$ with
$i=1,2,3,4$.  The position angle $\theta_p$ of the lens principal axis
is treated as a free variable.

The procedure is simple.  First
substitute the four observed image positions in eq.~\ref{rxy} 
to obtain the following four linear equations
\beq
\frac{\cos\theta'_i}{r_i} X_a +
\frac{\sin\theta'_i}{r_i} Y_a +
\frac{1}{r_i \cos\theta'_i} X_b -
\frac{1}{r_i \sin\theta'_i} Y_b = 1 ,~~~i=1,2,3,4,
\eeq
of the four unknown parameters $(X_a,Y_a,X_b,Y_b)$.  After solving these,
either analytically or numerically, 
these parameters are substituted in eqs.~\ref{XY},~\ref{xysp}, and~\ref{uvdef}
to yield the source position and the lens shape in terms of 
$(r_s,\theta_s,\delta,\beta)$.  In fact,
we can recast eq.~\ref{XY} to a set of four simple linear equations
of four new unknowns $(u,v,1/x'_s,1/y'_s)$ by
moving the terms $x'_s$ and $y'_s$ to the left hand side of the equations, 
i.e.,
\beq\label{XY2}
X_a/x'_s - v+u=1,~~~
Y_a/y'_s - v+u=1,~~~
X_b/x'_s - u=0,~~~
Y_b/y'_s - v=0.
\eeq
The lens shape parameters $(\delta,\beta)$ can then be computed from
\beq
\delta = \frac{u-v}{u+v},~~~\beta = \frac{1}{2(u-v)},
\eeq
and the source position $(r_s,\theta_s)$ from
\beq
r_s=\sqrt{{x'_s}^2+{y'_s}^2},~~~\theta_s=\arctan(y'_s,x'_s)+\theta_p.
\eeq
Thus we have effectively reduced the problem of fitting image positions
to successively solving linear equations, which is a straightforward 
task.

\subsection{Non-parametric radial profile and power-law slope}

The radial part of the potential can also be extracted from eq.~\ref{tr}
without parameterization.  At the positions of the images we have
\beq\label{radial}
m(\omega) = r^2-rr_s\cos(\theta-\theta_s)=x^2+y^2-xx_s-yy_s,~~~\omega=r f(\theta),
\eeq
where we have used $\partial_r \ln\omega=r^{-1}$.
Thus we have obtained a mass-radius relation directly from the observed image
positions, assuming that the source position $(x_s,y_s)$ and 
the flattening and position angle of the potential $(\delta,\theta_p)$
have been determined by fitting the curve (cf. eq.~\ref{rs}).
Interestingly, in the limit that the source is at the center of a
circular lens, we have $r_s\rightarrow 0$, $\beta \rightarrow 0$ and
$m \rightarrow r^2$.

It is useful to characterize the radial profile of a lens galaxy,
which is generally not a power-law, by an effective power-law slope 
$\alpha (\omega)$,
which varies with the radius $\omega$ except for scale-free models.
There are several ways of estimating the characteristic power-law slope.
Taking any two images $i$ and $j$, we can form a characteristic power-law slope
$\alpha_{ij}$
from the mass $m_i$ and $m_j$ at the image radii $\omega_i$ and $\omega_j$
with
\beq\label{alphamij}
\alpha_{ij} \equiv { \log m_i -\log m_j \over \log \omega_i - \log \omega_j }
={2 \log {r_i \over r_j} + \log {
1-r_s r_i^{-1} \cos(\theta_i-\theta_s) \over
1-r_s r_j^{-1} \cos(\theta_i-\theta_s) }
\over 
\log{r_i \over r_j} + \beta \log{ 
1-\delta\cos(\theta_i-\theta_s) \over 1-\delta\cos(\theta_j-\theta_s) }},
\eeq
where we have used the mass-radius eq.~\ref{radial}. 

Alternatively we can estimate the power-law slope from 
the observed time delay.  First
we rewrite the time delay eq.~\ref{tdelay}, so that
the time delay, $t_{ij}$, between any two images $i$ and $j$ is given by
\beq\label{tdelayij}
{h t_{ij} \over \tau_{100}} = 
{1 \over 2}(x_i^2+y_i^2)-
{1 \over 2}(x_j^2+y_j^2)-(x_i-x_j) x_s-(y_i-y_j) y_s-(\psi_i-\psi_j),
\eeq
where $\psi_i-\psi_j$
is the difference in the lens potential between the two images.
Rewriting eq.~\ref{tdelayij} we form a new estimator $\alpha^t_{ij}$ for 
the power-law slope with
\beq\label{alphatij}
\alpha^t_{ij} \equiv {m_i -m_j \over \psi_i- \psi_j}
={(x_i^2+y_i^2)-(x_j^2+y_j^2)-(x_i-x_j) x_s-(y_i-y_j) y_s
\over 
{1 \over 2}(x_i^2+y_i^2)-{1 \over 2}(x_j^2+y_j^2)-(x_i-x_j) x_s-(y_i-y_j) y_s-
{h t_{ij} \over \tau_{100}} },
\eeq
where we have used eq.~\ref{radial}.  

Thus we have two direct 
estimators $\alpha_{ij}$ and $\alpha^t_{ij}$  of the radial profile,
computed from the observed images and time delays.
In the limit of scale-free power-law models 
$\alpha_{ij}=\alpha^t_{ij}=\alpha(\omega)=cst$.
So the deviation from scale-freeness can be estimated 
by taking the differences such as
$\alpha_{12}-\alpha_{34}$,
$\alpha^t_{12}-\alpha^t_{34}$, or 
$\alpha_{14}-\alpha^t_{14}$.

\subsection{Hubble constant and time delay ratios}

To apply the above estimates of the power-law slopes, we have 
assumed that we know the rescaled Hubble constant $h$
from independent observations.  Alternatively
the Hubble constant $H_0=100h$ can be estimated 
from the time delay $t_{ij}$ between two images.
Letting $\alpha^t_{ij}=\alpha_{ij}$, we get
\beq\label{hij}
h = {\tau_{100} \over t_{ij}}
\left\{\left({1 \over 2}-{1 \over \alpha_{ij}}\right) 
\left[(x_i^2+y_i^2)-(x_j^2+y_j^2)\right]
-\left(1-{1 \over \alpha_{ij}}\right) 
\left[(x_i-x_j) x_s+(y_i-y_j) y_s\right]\right\}.
\eeq
The time delay ratio can also be predicted with
\beq\label{tratio}
{t_{ij} \over t_{kl}} = {
\left({1 \over 2}-{1 \over \alpha_{ij}}\right) \left[(x_i^2+y_i^2)-(x_j^2+y_j^2)\right]
-\left(1-{1 \over \alpha_{ij}}\right) \left[(x_i-x_j) x_s+(y_i-y_j) y_s\right],
\over
\left({1 \over 2}-{1 \over \alpha_{kl}}\right) \left[(x_k^2+y_k^2)-(x_l^2+y_l^2)\right]
-\left(1-{1 \over \alpha_{kl}}\right) \left[(x_k-x_l) x_s+(y_k-y_l) y_s\right] },
\eeq
which is obtained by rewriting eq.~\ref{hij}.

\section{Application: the surface density and time delay models for PG1115+080}

\subsection{Data}

As a simple application, we model the image positions and time delays
of the well-studied quadruple system PG1115+080.  This system has been
extensively studied ever since the first models by Young and
collaborators (1981), and has received closer attention after
Schechter et al.'s (1997) measurements of its time delay.  All
models, except those of Saha \& Williams (1997), adopt
elliptical/circular shapes and a few common radial profiles, with the
models of Keeton \& Kochanek (1997) and Impey et al. (1998) being the
most comprehensive.  Only one year after the discovery of the
legendary double-image radio-loud quasar Q0957+561, this system was
identified as a 
multiple-imaged system by Weymann et al. (1980) in their survey of
nearby bright QSOs.  It is now known to consist of four images with
the names $A_1$, $A_2$, $B$ and $C$ (with flux ratios about 4 : 2.5 :
0.7 : 1, cf. the HST observations of Kristian et al. 1993) of a
radio-quiet QSO at redshift $z_s=1.722$.  The images $A_1$ and $A_2$
are within $0.5''$; see the inset of Fig.~\ref{geometry}.
Interestingly, the lens galaxy is also one of the bright members of a
galaxy group ($N \sim 10$) at redshift $z_l=0.310$, first mapped by
Young et al. (1981).  The center of the group is to the south-west of
the lens, roughly at $r_g=(20''\pm 2'')$ and $\theta_g=(-117^o \pm
3^o)$.  The lens galaxy has been resolved by both HST and the 8.2-m
Subaru telescope in $0.3''$ seeing.  It appears to be an early type
galaxy with a de Vaucouleurs profile and a half-light radius of
$0.55''$.  There is no sign of differential dust-extinction in the
lens galaxy.  While NICMOS observations by Impey et al. (1998) show no
flattening for the lens, ground infrared images by Iwamuro et
al. (2000) found it to be an E1 galaxy elongated towards $\theta_p
\sim 65^o$.  Both observations reveal a $1''$ infrared Einstein ring
connecting the four images, which is thought to be the infrared image
of the QSO host galaxy.  PG1115+080 is also one of the two quadruple
systems where the time delay between images has been measured, the
other one being the radio-loud quasar B1608+656 from the CLASS
survey (cf. Fassnacht et al. 1999).  Although 
two different sets of values are quoted in the literature (Schechter
et al. 1997, Barkana 1997), the leading image is the furtherest image
(the image C), and the innermost image (image B) arrives last.  The
time delay ratio $r_{ABC}=t_{AC}/t_{BA}$, for the delay between image
$C$ and $A_1+A_2$ vs. image $B$ and $A_1+A_2$, provides an extra
discriminator of the models; the images $A_1$ and $A_2$ are within
$0.5''$ of each other, and the small relative delay is
undetected. Schechter et al. first reported $r_{ABC} = 0.7\pm 0.3$
from their photometric monitoring program in 1995-1996.  Later
analysis by Barkana (1997) found $r_{ABC}=1.13 \pm 0.18$, after taking
into account correlations of errors amongst the time delays. The delay
between images B and C, $t_{BC}=25.0 \pm 1.7$ days.

Here we illustrate the application of our models to 
the most recent data from Impey et al. (1998) of PG1115+080.  We do
not attempt to model B1608+656 because the lens appears to be a
merging pair of galaxies, and the morphology is too complex for our
model.  We denote with the index $i=1,2,3,4$ the four images 
$A_1$, $A_2$, $B$ and $C$.  
All results are quoted for a standard flat universe without
a cosmological constant $(\Omega, \Lambda)=(1,0)$.  
Table~\ref{cosmotab} gives the relevant quantities to calibrate
our results to other universes.
The angular distance from redshift $z_1$ to $z_2$ in a universe
of a matter and vacuum density
$(\Omega,\Lambda)$ times the closure density is generally given by
\beq
D(z_1,z_2) = {c \over H_0 (1+z_2) \sqrt{\Omega_c}} 
\sinh\left\{ \sqrt{\Omega_c} \int_{z_1}^{z_2}dz
\left[\Lambda + \Omega_k (1+z)^2 + (1+z)^3\Omega \right]^{-{1 \over 2}}\right\},
~~\Omega_k=1-\Lambda-\Omega.
\eeq
The predicted Hubble constant should be reduced from the value
achieved with the standard $(\Omega,\Lambda)=(1,0)$ universe by 3\% in
the presently favored $\Lambda$-dominated universe with
$(\Omega,\Lambda)=(0.3,0.7)$ from surveys of distant supernovae.

\subsection{The source position, the lens shape and mass inside images}

First we solve for the lens shape and source position from the linear equations
~\ref{rxy} and~\ref{XY2}.  
The solutions for PG1115+080 are given in Table~\ref{betatab}, sorted
according to the value of $\beta$ or $\theta_p$.
The resulting potential model with $\beta=-{1 \over 8}$
or $\theta_p=60.5^\circ$ has
a flattening of between E0 and E1, and 
interestingly the lens principal axis points towards
the location of the galaxy group in the lens plane.  Models
with other values of $\beta$ or $\theta_p$ 
will be discussed in section 5.

To proceed with determining the lens mass at each image position, 
we substitute the now
known flattening parameters $(\delta,\theta_p)$ and the source
positions $(r_s,\theta_s)$ in eq.~\ref{radial}, to predict four
independent data points $(\omega_i,m_i)$ with $i=1,2,3,4$ in the
radius vs. the enclosed mass plane.  
Fig.~\ref{massrad} shows the predicted lens mass enclosed at the four image
positions.  Note that the mass rises faster than the light, implying a
growing dark mass component at large radius; the light
distribution is modeled as an observed de Vaucouleurs $r^{{1 \over 4}}$-law 
with a half-light radius of $0.55''$.

\subsection{Piecewise power-law model}

So far we did not enforce any strict parameterization of the radial
profile. We only restrict the profile to be of the form of
eq.~\ref{omegadef}; in practice, this is an insignificant
restriction. In the following sections we show several ways of
modeling the radial profile assuming an isolated lens.  None of the
models is completely satisfactory.

First we use a minimal model, which assumes
that the mass-radius relation is a piecewise power-law, that is,
we connect a straight line through two images in the $\log m$
vs. $\log \omega$ plane.  The piecewise values for the power-law slope
and axis ratios are given in Table~\ref{betatab}.  

The Hubble constant $H_0=100h$ can be estimated 
by normalizing the time delay to
the observed value $t_{AC}$ (Schechter et al. 1997), 
\beq\label{hAC}
h = {\tau_{100} \over 2t_{AC}} 
\left[-P_{14}+\left(1-{1 \over \alpha_{14}}\right) S_{14}\right],
\eeq
where
\beq
P_{ij}=\left[(x_i^2+y_i^2)-(x_j^2+y_j^2)\right]
\eeq
depends on the image positions $(x_i,y_i)$ alone and
\beq
S_{ij}=2P_{ij}-2\left[(x_i-x_j) x_s+(y_i-y_j) y_s\right]
\eeq
depends on the source position $(x_s,y_s)$ as well.  The model yields a
Hubble constant $H_0 \sim 30$~km/s/Mpc, much lower than determined by
other authors (e.g., Impey et al. 1998).  The low $H_0$ is a result
of the high value for the power-law slope $\alpha_{AC}=1.6$.
Other values are given in Table~\ref{betatab}
for various image pairs and observed time delay.
The time delay ratio can also be predicted with (cf. eq.~\ref{tratio})
\beq\label{rabc}
r_{ABC} \equiv {t_{14} \over t_{32}} = {
- P_{14}
+\left(1-{1 \over \alpha_{14}}\right) S_{14}
\over
- P_{32}
+\left(1-{1 \over \alpha_{32}}\right) S_{32}
}.
\eeq
Substituting the slopes $\alpha_{AC}$ and $\alpha_{BA}$ to 
eq.~\ref{tratio} we find the ratio between images $r_{ABC}=t_{AC}/t_{BA} \sim
0.65$.

Note that
the time delay predictions here are robust and independent of details
of the density, 
since the discontinuity in the density is completely smoothed out in 
the lensing potential.  

\subsection{Single power-law model}

\begin{figure}
\vskip -0.5cm
\epsfxsize=23.5cm
\centerline{\epsfbox{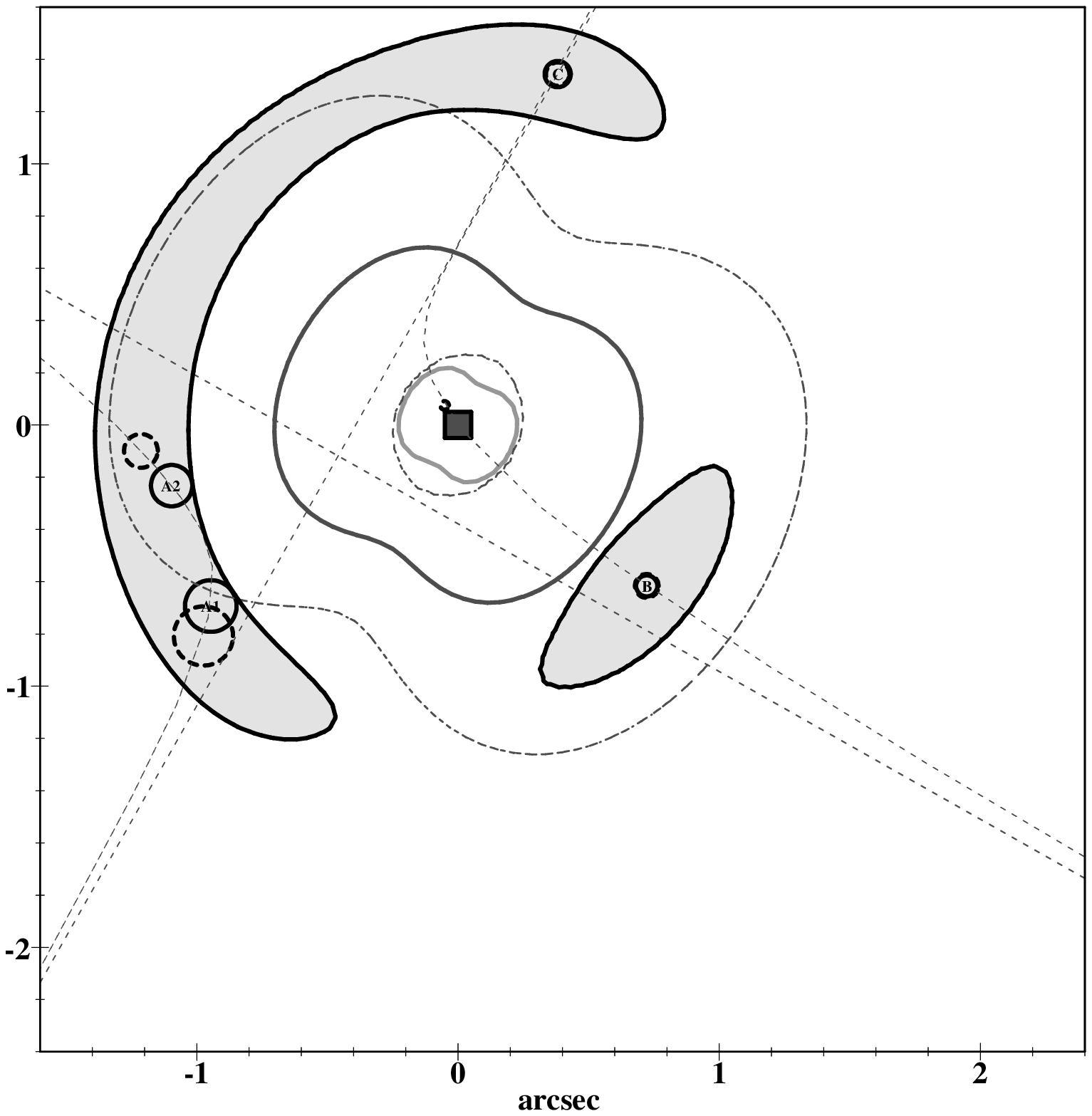}}
\vskip -11cm
\epsfysize=10cm
\rightline{\epsfbox{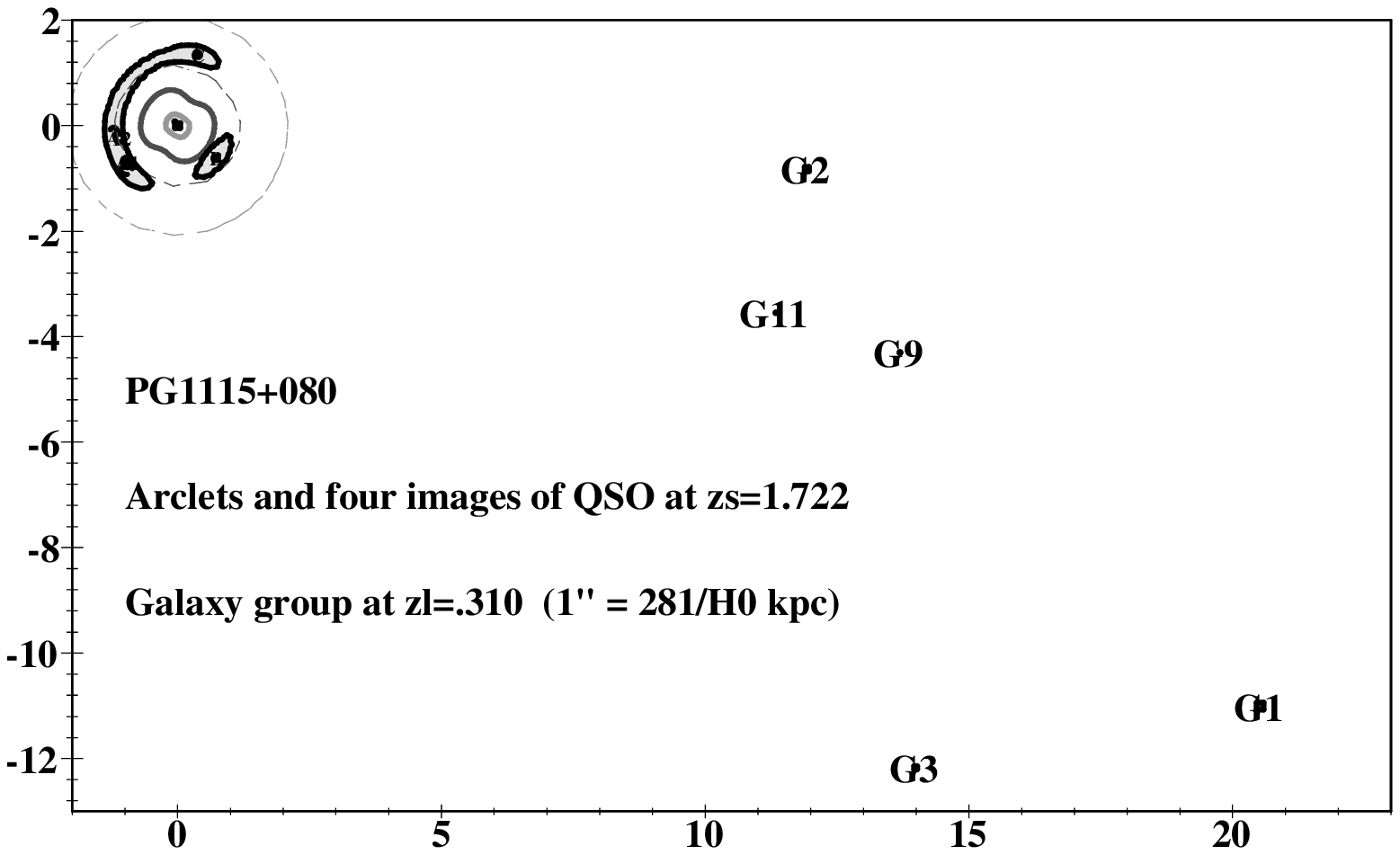}}
\vskip -1cm
\caption{
shows the relative locations of the four QSO images and the lens galaxy
(filled square at the center) in the PG1115+080 system.
The inset shows the relative locations of PG1115+080 and 
the neighbouring galaxies (G1 to G11 in the notation of Impey et al. 1998). 
The equation~\ref{curve} for self-similar models 
also predicts a semi-hyperbolic curve 
passing through the observed images (solid circles), the predicted images
(dashed circles) and the source (smaller half-closed circle) and the lens (the
filled square) with asymptotic lines always 
being parallel to the symmetry axes of the potential.  
The area of the circles is in proportion
to the flux.  Overplotted are 
the predicted surface density contour maps (solid contours
for density being 1 and 2 times the critical density) 
and critical curves (dashed contours)  in the single-power-law model with 
$\beta=-{1 \over 8}$.
The nearly closed ring is predicted for a uniform disk around the QSO.
}\label{surf2}\label{geometry}
\end{figure}

The piecewise-power-law model above necessarily implies a
discontinuous density profile.  This could be cured if we enforce a
single power-law model, that is, we fit a straight line to the four
points in the $\log m$-$\log \omega$ plane.  This would give us a
power-law slope $\alpha=1.38$.  Fig.~\ref{surf2} shows our model
surface density contours.  Similar to the non-parametric models of
Saha \& Williams (1997) we find a peanut-shaped lens.

We can estimate the goodness of the fit by recomputing the image
positions from a given mass model.  The general procedure 
of simulating images of our theoretical lens model is as follows:
First combine eq.~\ref{radial} with 
the power-law radial profile to get
\beq
r^2-rr_s\cos(\theta-\theta_s) = m(\omega) 
=b_0 r_0 \left({\omega \over r_0}\right)^\alpha,
~~~
\omega=r\left| 1-\delta \cos (2\theta-2\theta_p)\right|^\beta.
\eeq 
Then upon substitution of eq.~\ref{curve} to eliminate $r$, we obtain
a one-dimensional non-linear equation for the image position angle
$\theta$, which can be solved easily numerically.  
As a comparison, one would be dealing with a minimum-finding or a
root-finding numerical problem in a two-dimensional plane $(r,\theta)$
in the general case without the separation of the angular vs. radial part.
For our model of PG1115+080, the image solutions are shown in
Fig.~\ref{surf2} as dashed circles, together with the
input observed image positions (solid circles). 
The predicted four images are off by about 60
milli arcsecs, a residual which is inconsistent with the $\sim 3$
milli arcsecs accuracy of Impey et al. positions, and is marginally
consistent with earlier data by Kristian et al. with a quoted error of
50 milli arcsec for the lens galaxy and 5 milli arcsec for the images.
The images $A_1$ and $A_2$ are at two sides of the critical curve (the
line of infinite amplification in the source plane), hence are highly
amplified with opposite parity.  The time delay ratio can be estimated
with eq.~\ref{tratio}, assuming a constant power-law slope
$\alpha=\alpha_{BC}=1.38$.  This yields $r_{ABC} \sim 1.3$, close to
the Barkana (1997) value of $1.13 \pm 0.18$.  The large difference
here is due to the large residual in terms of fitting the images $A_1$
and $A_2$ with a straight power-law.

We also compute the amplification patterns by taking double
derivatives of the time delay surface.  The circles in Fig.~\ref{surf2} 
show the observed (solid circles) and predicted
(dashed circles) fluxes of each image and the source (half-closed circle), 
with the area of
each circle in proportion to the flux.  The $B$ to $C$ ratio is well-reproduced
and the predictions for images $A_1$ and $A_2$ are also consistent
with observations at about the 0.3 magnitude level.  

Finally, if we put a host galaxy around the QSO, the model predicts
that the image of the host galaxy will be stretched into an arc.  We
see a nearly closed ring.  This agrees very well with the diffuse ring that
Impey et al. discovered in their NICMOS images.  The ring maps
back to the source plane as a disk with an area of $\sim 0.03$ square
arcsec.

\subsection{Double power-law model}

\begin{figure}
\epsfysize=20cm 
\centerline{\epsfbox{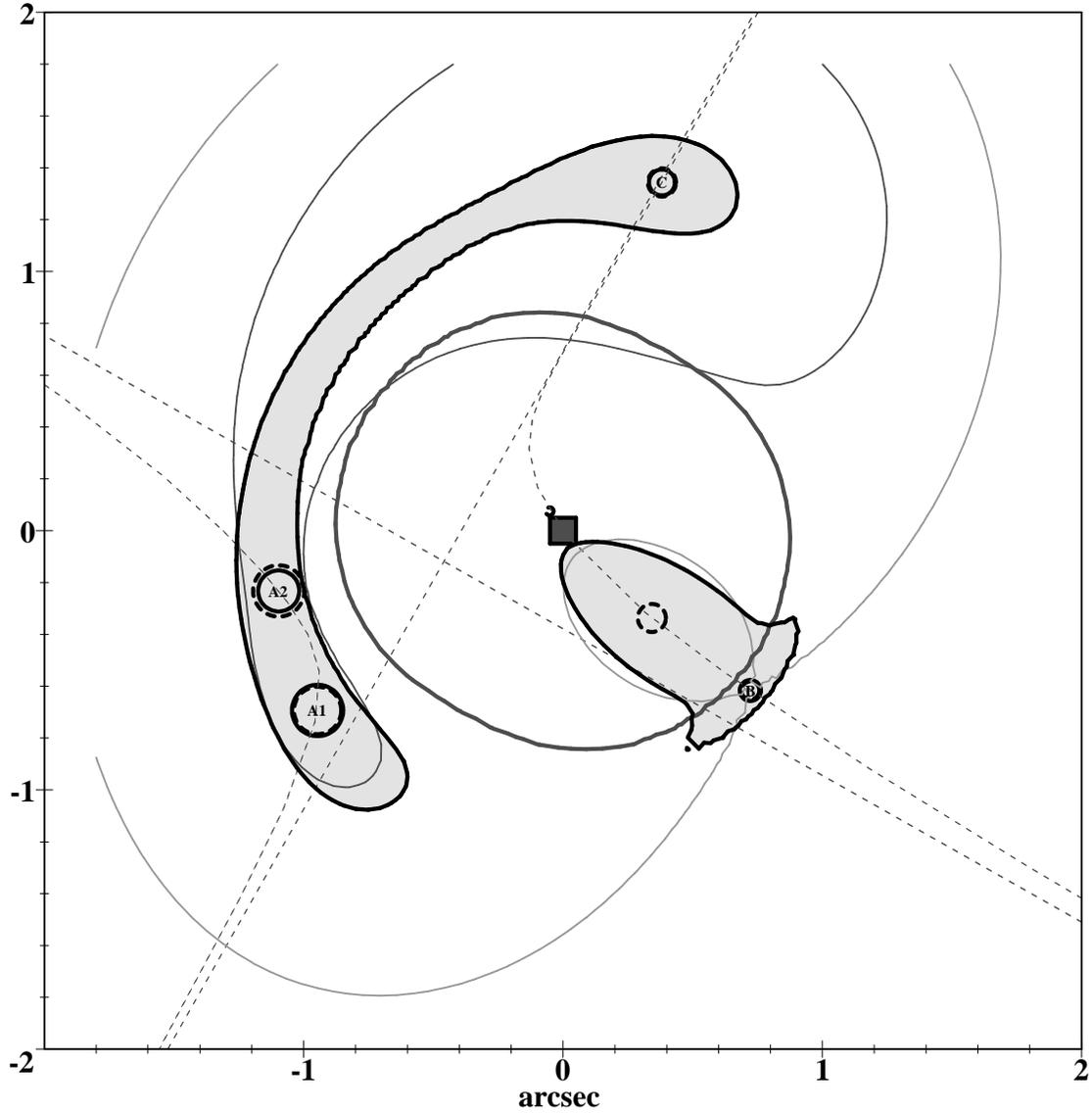}}
\caption{
Similar to Fig.~\protect{\ref{surf2}}, but for the double-power-law model
with $\beta=-{1 \over 8}$.
Overplotted is a solid contour, showing the 
predicted surface density at the critical density.
Also shown is 
the predicted time delay ``finger print'' for the double-power-law model with 
$\beta=-{1\over 8}$.
The contour passing close to the separatrix (image $B$) 
is split to two disconnected contours.   
The extra fifth image (the isolated dashed circle near the image B) 
is a result of the non-monotonic model density, 
and it sits on a local extreme of the delay contour.
The other contour is a contour of constant 
time delay ratio $t_{XC}/t_{BX}=0.7$ of any point $X$ 
and images $B$, $C$.  
}\label{surf}\label{td}
\end{figure}

Alternatively we can fit a smooth, five-parameter 
lens model with a lens potential
\beq\label{double}
\psi(\omega)= c_0
\left({\omega \over a_0}\right)^{\alpha_{in}}
\left[1+\left({\omega \over a_0}\right)^n\right]^{\alpha_{out}-\alpha_{in} \over n},
~~~\omega=r\left| 1-\delta \cos (2\theta-2\theta_p)\right|^\beta.
\eeq
The corresponding mass profile is given in Appendix B.  
This model assumes the lens potential 
(as well as the lens mass) increases like a double-power-law
with an inner slope $\alpha_{in}$ and outer slope $\alpha_{out}$.
The transition is defined by the normalization constant $c_0$, 
the radius $r=a_0$ and the sharpness parameter $n$; 
bigger $n$ corresponds to sharper transition. When fitting the mass model 
to the four points in Fig.~\ref{massrad},
it turns out that the inner slope $\alpha_{in}$ is fully unconstrained.
The other four free parameters $(c_0,a_0,\alpha_{out},n)$ are determined 
from fitting the four data points; the procedure is explained in Appendix B.
The values of $(c_0,a_0,\alpha_{out}) \sim (0.6,1.1,1.63)$, 
nearly independent of the value for the inner slope $\alpha_{in}$.
Note $\alpha_{out} \sim \alpha_{AC}$, consistent with the fact that
the mass at
$A_1$, $A_2$ and $C$ follows nearly a power-law (cf. Fig.~\ref{massrad}).
The value of $n$ increases from $\sim 15$ to $\sim 39$ 
for $\alpha_{in}=0-2$.  
That $n \gg 2$ means that the density changes sharply at the transition.
All fits have zero residual and 
predict nearly identical mass profiles at radii
between images $B$ and $C$.  They differ only
in the mass profile inside image $B$, where we have no direct
constraint on the dark matter profile.  For all purposes it is sufficient 
to set $n=20$ so that $\alpha_{in} \sim 2$, in which case
the model has a finite core at small radii.  
From the potential model we can compute
the dimensionless surface density contours (cf. Fig.~\ref{surf}).
Near the position of the images, the density has a flattening of E2-E3,
flatter than the potential, as expected.  The potential model
can also be substituted in eq.~\ref{tdelay} to predict the time delay
contours, shown in Fig.~\ref{td}.  
The observed images $C$, $A_2$, $A_1$, $B$ (the solid
circles) are exactly the valley, peak, saddle, valley points of
the delay contour, where the theoretical images should lie 
according to Fermat's principle.  The images $A_1$ and $A_2$
have nearly the same arrival time.  
Substituting the lens potential in eq.~\ref{tdelayij} 
we can predict the time delay ratio for 
the double-power-law model.  The result, $r_{ABC} \sim 0.65$, 
is nearly the same as the piece-wise power-law model.  
These predicted ratios are in good agreement with 
the earlier measurement of $0.7 \pm 0.3$ of Schechter et al. (1997).

We can recompute the image
positions from the double-power-law model by solving 
\beq
r^2-rr_s\cos(\theta-\theta_s) = m(\omega) 
=c_0 \left({\omega \over a_0}\right)^{\alpha_{in}}
\left[1+\left({\omega \over a_0}\right)^n\right]^{\alpha_{out}-\alpha_{in} \over n},
~~~
\omega=r\left| 1-\delta \cos (2\theta-2\theta_p)\right|^\beta,
\eeq 
and eq.~\ref{curve}.  

Unlike the single-power-law model it predicts five images
(cf. Fig.~\ref{surf}). Four predicted images fall exactly on the
observed positions. But the model predicts an extra image near 
image B. This can be identified with the usually highly demagnified
fifth image that is sometimes observed in lensed systems. The extra
image the model predicts is, however, too bright to be consistent with
observations. This is a direct consequence of the fact that $\kappa$
at its radius is nearly constant and $\sim 1$. The amplification can
be calculated from $\kappa$ as $(1-\kappa)^{-2}\sim25$.

\begin{figure}
\epsfysize=20cm 
\epsfxsize=20cm 
\centerline{\epsfbox{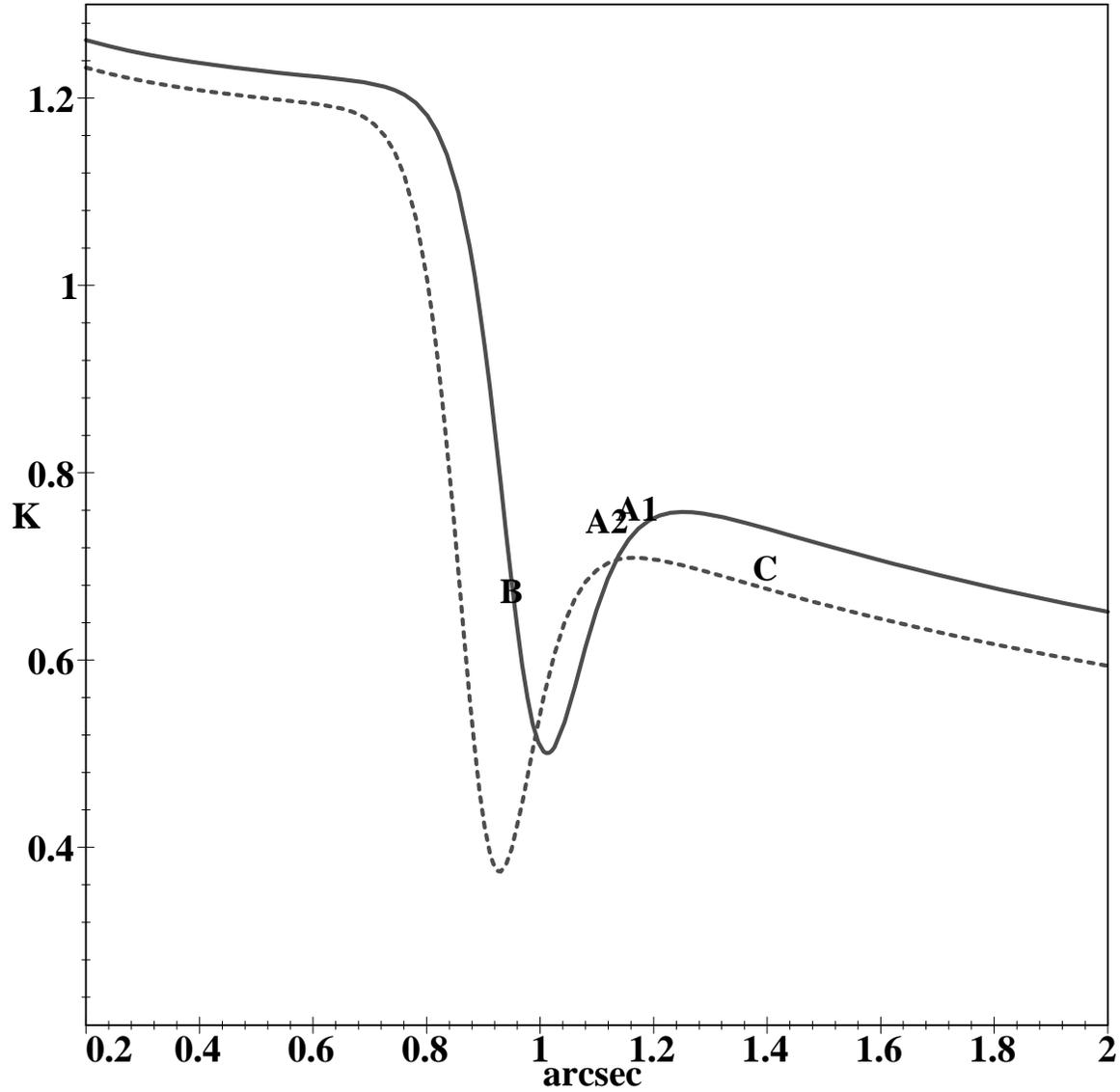}}
\caption{
The predicted run of the dimensionless density $\kappa$ along (solid) and 
perpendicular to (dashed) the principal axis $\theta_p=60.5^o$
(i.e. $\beta=-{1 \over 8}$) 
for the double-power-law model.  The model is positive everywhere, but
has a peculiar wiggle at about $1''$.
}\label{denprofile}
\end{figure}

All images lie on the curve defined by eq.~\ref{curve}.  The extra
image arises because the model density profile
(cf. Fig.~\ref{denprofile}) is non-monotonic near $1''$ radius, the
transition radius $a_0$ of the double-power-law potential model.  A
wiggle in density happens when the transition parameter $n>2$, i.e., a
sharp transition of the potential.

\section{Discussion}

\subsection{Power-law slope of PG1115+080 vs. $H_0$ and time delay ratio}

\begin{figure}
\epsfysize=10cm 
\centerline{\epsfbox{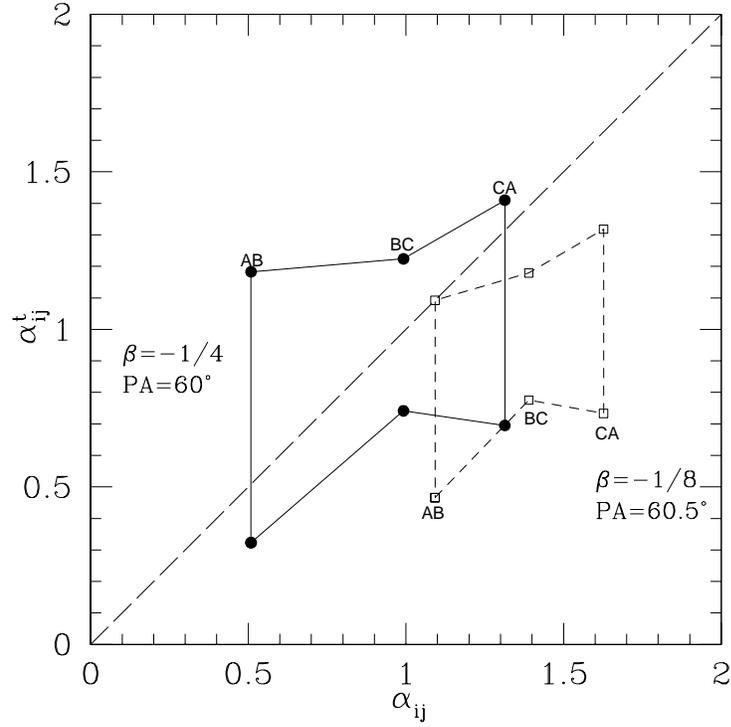}}
\caption{
The predicted power-law slope of the lens galaxy, $\alpha_{ij}$
from fitting the mass-radius relation (the horizontal axis, 
cf. eq.~\protect{\ref{alphamij}}), 
and $\alpha^t_{ij}$
from fitting the time delay (the vertical axis, 
cf. eq.~\protect{\ref{alphatij}}), where
the $\pm 2\sigma$ range of Barkana time delay and $50<H_0<100$ are used
with higher $H_0$ corresponding to lower $\alpha^t_{ij}$.
The predictions are made for fitting the pairs
$(A_1, C)$, $(B, C)$, and $(A_2, B)$.
}\label{alpha}
\end{figure}

The previous sections have illustrated the procedure for fitting known
quadruple image systems with our models, and have also
revealed some puzzling problems with PG1115+080, for example, the low
and non-unique $H_0$ and the peculiar and non-unique lens density.
Fig.~\ref{alpha} shows the values for $\alpha$ predicted from either
the mass-radius relation (cf. eq.~\ref{alphamij}) or from the time
delay (cf. eq.~\ref{alphatij}).  Two comments are in order: (1) There
is a large spread with the power-law slope $\alpha$ depending on how
it is predicted. To fit the observed
time delay within $2\sigma$ with a reasonable $H_0$ (between 50 and
100 km/s/Mpc), we obtain only a loose constraint $0.3< \alpha <1.6$.
(2) The rise of the power-law slope $\alpha$ with radius 
in the piecewise-power-law model is somewhat unphysical; realistic
models typically have a monotonically steeper 
density profile (smaller $\alpha$) with increasing radius.  These results
are consistent with the finding of a wiggle near the transition radius 
($1''$) in the surface density of the smooth double-power-law fits.  
Nevertheless the density is positive everywhere.  

A detailed treatment of these problems should include the effect
of the neighboring group.  Such models are presented elsewhere, since
they are beyond the interest of this paper on methods for quadruple
systems in general.  Nevertheless some further study of the model
parameter space is clearly needed.  Here it is sufficient to comment
on the range of the predicted model parameters, in particular, the
power-law slope and the value of $H_0$ when we vary the parameter
$\beta$ or $\theta_p$.

\begin{figure}
\epsfysize=20cm 
\centerline{\epsfbox{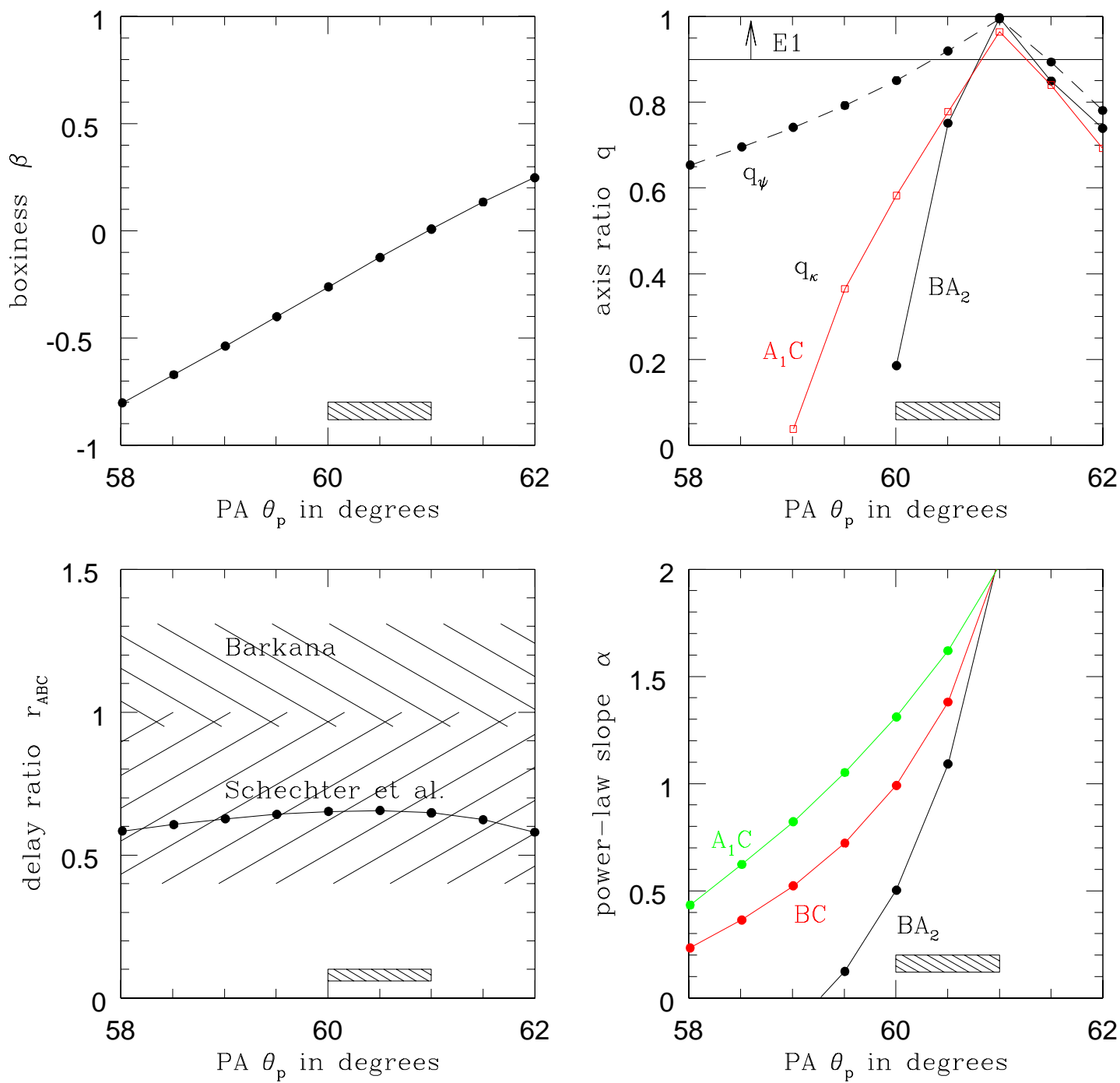}}
\caption{
The parameter space of the isolated lens model.  
The four panels show (in clockwise direction) the model boxiness
parameter $\beta$, the axis ratio $q_\psi$ of the potential 
and $q_\kappa$ of the density (cf. eq.~\ref{qpsi} and eq.~\ref{qkappa}), 
the power-law slope $\alpha$ (cf. eq.~\ref{alphamij}), 
and the delay ratio $r_{ABC}$ (cf. eq.~\ref{tratio}). 
The shaded bar at the bottom of each panel indicates the
range of physically allowed models.
}\label{paraspace}
\end{figure}

\begin{figure}
\epsfysize=10cm 
\centerline{\epsfbox{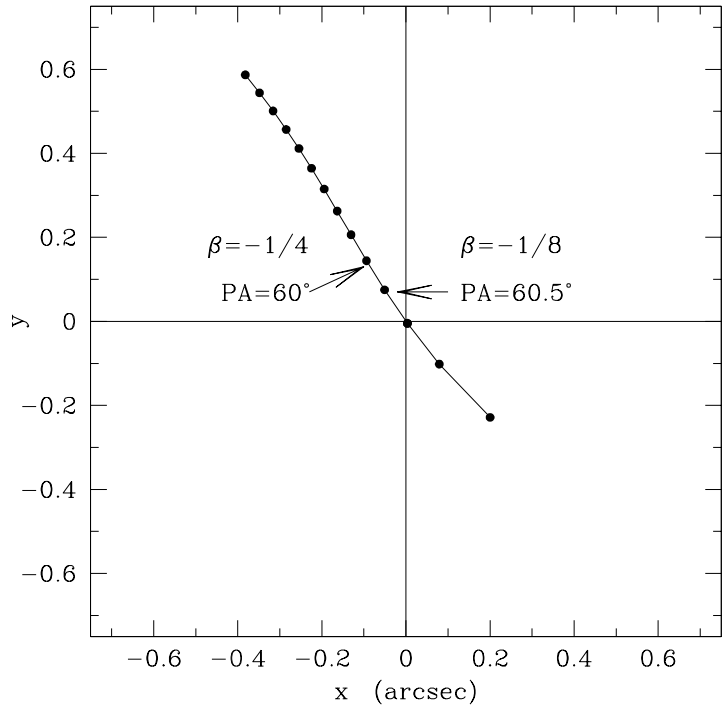}}
\caption{the predicted source position for the sequence of models with the
principal axis in the range $58^o-62^o$ (from the upper left
to the lower right), which corresponds to $\beta=-0.8$ to $\beta=0.25$.  
The source lies on a nearly straight line passing the origin, where
the lens is.
}
\label{sourcepos}
\end{figure}

Fig.~\ref{paraspace} shows the parameter space of the isolated lens
model by varying the orientation of the lens principal axis $\theta_p$
(cf. eq.~\ref{ftheta} and \ref{thetaprime}).
It turns out a one-parameter family of models fits the observations;
we effectively vary $\beta$.  In fact,
$\theta_p \sim 61^o + 4\beta$, and $\Delta =2\beta \delta \sim 
{\beta \over 2}$ to a fair accuracy.
We find that models with $\beta <-1/3$ will generally result
in negative density models, models with $\beta \ge 0$ have a
radially-increasing density, and models with $|\beta| \rightarrow 0$ also
produce a nearly constant axisymmetric density, hence implying an
unphysically small Hubble constant.  Only in the range $-{1 \over 4}
\le \beta < 0$ do we find physical models with a reasonable
flattening.  
The source positions for different models are shown in Fig.~\ref{sourcepos}.
Table~\ref{betatab} compares the predicted parameters of the
$\beta=-{1 \over 4}$ or $\theta_p=60^\circ$ model 
with the $\beta=-{1 \over 8}$ or $\theta_p=60.5^\circ$ model.
Fig.~\ref{massrad} shows the predicted radial profile for both models.
Both models would have a significant residual if we fit the image
positions with a straight power-law, or strange wiggles and extra
images if we fit a double-power-law.
We find this is generally the case with all isolated
lens models without the shear from the neighbouring group.

Fig.~\ref{hgraph} shows that there is significant degeneracy
with the value for $H_0$, depending on the adopted model and the adopted
time delay.  The most reliable estimate is from the delay between the 
innermost image B and the outermost image C using
\beq\label{hBC}
h = {\tau_{100} \over 2t_{BC}} 
\left[
-P_{14}+\left(1-{1 \over \alpha_{14}}\right) S_{14} 
-P_{32}+\left(1-{1 \over \alpha_{32}}\right) S_{32}\right],
\eeq
where we neglected the time delay between the images $A_1$ and $A_2$.
The $H_0$ thus predicted scales approximately as $60
(2-\alpha)$~km/s/Mpc (cf. Fig.~\ref{hgraph}), where
$\alpha=\alpha_{BC}$ is the average power-law slope.
The factor
$2-\alpha$ is the average power-law slope of the model surface density, and
is not well-constrained at the radius of the images.  
Models with $\beta \sim -{1 \over 4}$ 
or $\theta_p=60^\circ$ 
predict a steeper density profile, and closer to isothermal 
than models with $\beta \sim -{1 \over 8}$ or $\theta_p=60.5^\circ$, 
hence yield a more plausible $H_0$ around $60$~km/s/Mpc.  Models with
very shallow profiles 
($\alpha>1$) are unfavored by the consensus value for $H_0$.  

In all our
isolated lens models, the delay ratio is closer to the Schechter value,
while the Barkana value seems to be the widely accepted one.
This explains why we see in Fig.~\ref{hgraph} 
a tighter prediction of $H_0$ using $t_{AC}$ and $t_{BA}$ of Schechter
values than the Barkana values.
Interestingly in the limit that the isothermal model is applicable,
$\alpha =1$, the time delay ratio can be estimated 
from the image positions directly (i.e. without computing the source position)
with the following equation:
\beq
r_{ABC} = {P_{14} \over P_{32}} 
={r_1^2-r_4^2 \over r_3^2-r_2^2} \sim 0.65,~~~\mbox{\rm for isothermal lens},
\eeq
where $r_i$ is the observed radius of the image $i$ from the lens center.
The ratio increases when we introduce the shear from the nearby group.

\subsection{Rotation curve and mass-to-light ratio}

\begin{figure}
\epsfysize=10cm 
\centerline{\epsfbox{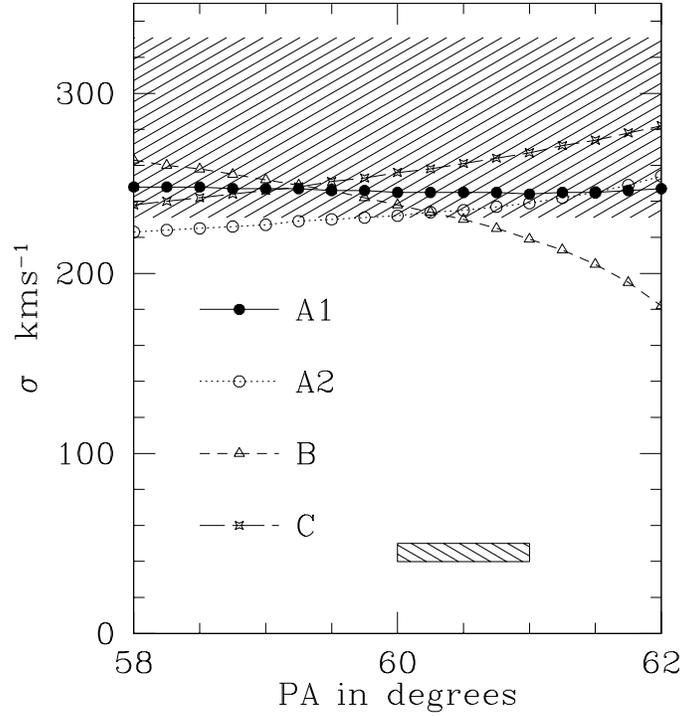}}
\caption{
The predicted velocity dispersion of dark matter in the lens galaxy
for a series of lens models specified by different
lens position angle $\theta_p$.
The values are calculated from the image positions of
$A_1$ (solid circle), $A_2$ (open circle), $B$ (triangle) and $C$ 
(starry symbol) respectively.  The shaded region shows the 95\% confidence
range from the observed average dispersion of stars in the lens (Tonry 1998).
}\label{sigpara}
\end{figure}

Fig.~\ref{sigpara} compares the predicted velocity dispersion for
the lens with the observed value.
The dispersion is predicted with the following formula,
\beq\label{sigma}
\sigma_i^2 = V^2 {m_i \over 2 r_i },
~~~V^2 \equiv {D_{os} \over D_{ls}} 
{(3\times 10^5{\rm km/s})^2 \over 2 \pi ~{\rm radian} } 
            {1 ~{\rm radian} \over 206265 ~{\rm arcsec}},
\eeq
where the predictions are made from each image $i=1,2,3,4$
(strictly speaking the formula is valid only for a singular isothermal lens).
Applying this for each value of the lens position angle $\theta_p$
we get the full range of possible values for the dispersion.
The difference of the dispersion among the four images shows the deviation
from an isothermal model.
On average the predicted lens galaxy velocity dispersion, 
200-300 km/s, suggests that the lens is close to a massive $L_*$ galaxy.
The observed dispersion $\sigma=(281 \pm 50)$ km/s ($95\%$ confidence,
shaded region) from the Keck spectrum of Tonry (1998).
This value is comparable to the theoretical models, 
but on the high side.
A similar problem has been noted in previous models (Schechter et al. 1997).
Tonry noted that the discrepancy might be due to a steep radial fall-off of the
velocity dispersion or a simple over-estimation of the observed dispersion;
the spectrum of the faint lens was made from subtracting two Keck spectra
taken with a $1''$-slit, one passing
the lens and the image B, one passing the images $A_1$ and $A_2$.

\begin{figure}
\epsfysize=10cm 
\centerline{\epsfbox{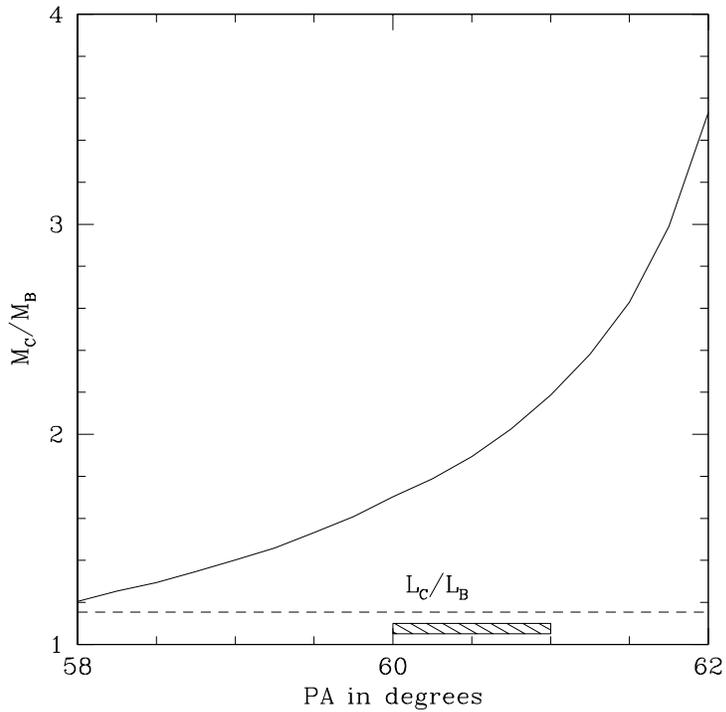}}
\caption{
Similar to ~\protect{\ref{sigpara}}, but for
the predicted ratio of the lens mass inside the 
outermost image $C$ and the innermost image $B$, $M_C/M_B$.  Note that 
the ratio is always higher than the ratio of the light inside the two images,
$L_C/L_B$ (the dashed line); 
the latter is calculated from the observed de Vaucouleurs law.
}\label{mratio}
\end{figure}

Fig.~\ref{mratio} compares the increase of the lens mass 
from the innermost image $B$ to the outermost image $C$ 
vs the increase of the lens light.  The mass ratio is
predicted using
\beq
{M_C \over M_B} ={m_4 \over m_3}=\left({r_4 \over r_3}\right)^{\alpha_{BC}}.
\eeq
where the power-law slope $\alpha_{BC}$ depends the principal axis $\theta_p$.
The light ratio
\beq
{L_C \over L_B} = {
\int_0^{r_4} dr 2\pi r\exp\left[-7.67\left({r \over r_e}\right)^{1 \over 4}\right]
\over 
\int_0^{r_3} dr 2\pi r\exp\left[-7.67\left({r \over r_e}\right)^{1 \over 4}\right]
},~~~r_e=0.55'',
\eeq
where the observed de Vaucouleurs law is used for the projected light.
The fact that ${M_C \over M_B} > {L_C \over L_B} >1$ implies
that the mass grows faster than the light at these radii for all these models,
consistent with a large amount of dark matter between 0.8 and 1.5 arcsec.

\section{Conclusion}

Here we itemize our main results.

(i) We found a very general class of lens models that allow for
   non-elliptical and non-scale-free lenses.  These models include
   previous isothermal elliptical models as special case.  We can
   derive the radial mass distribution of the lens in a non-parametric
   way.  We can study the deviation from the usually assumed straight
   power-law profile.  We also give simple formulas for computing the
   time delay ratios and for estimating $H_0$.

(ii) The models are very easy to compute and can be used for efficient
   exploration of a large parameter space because the lens equations
   can be reduced to a set of linear equations.  

(iii) We have applied the models to PG1115+080, and have explored 
   a large parameter space of isolated lens models.  
   All models using piece-wise-power-law or
   double-power-law fit the positions of the images exactly.
   The models can also reproduce the flux ratios between the images and the
   stellar velocity dispersion of the lens approximately.  

(iv) Our models are consistent with a dark halo up to a radius of
   three times the half-light-radius of the lens.  The enclosed mass   
   increases much faster than the enclosed light as we move radially
   from the innermost image
   to the outermost image.

(v) We reconfirm earlier results by (e.g. Schechter et al. 1997) that
   the principal axis of the lens potential points, within a few
   degrees, to the external group, and is consistent with the observed
   value (Iwamuro et al. 2000).

\begin{figure}
\epsfysize=20cm 
\centerline{\epsfbox{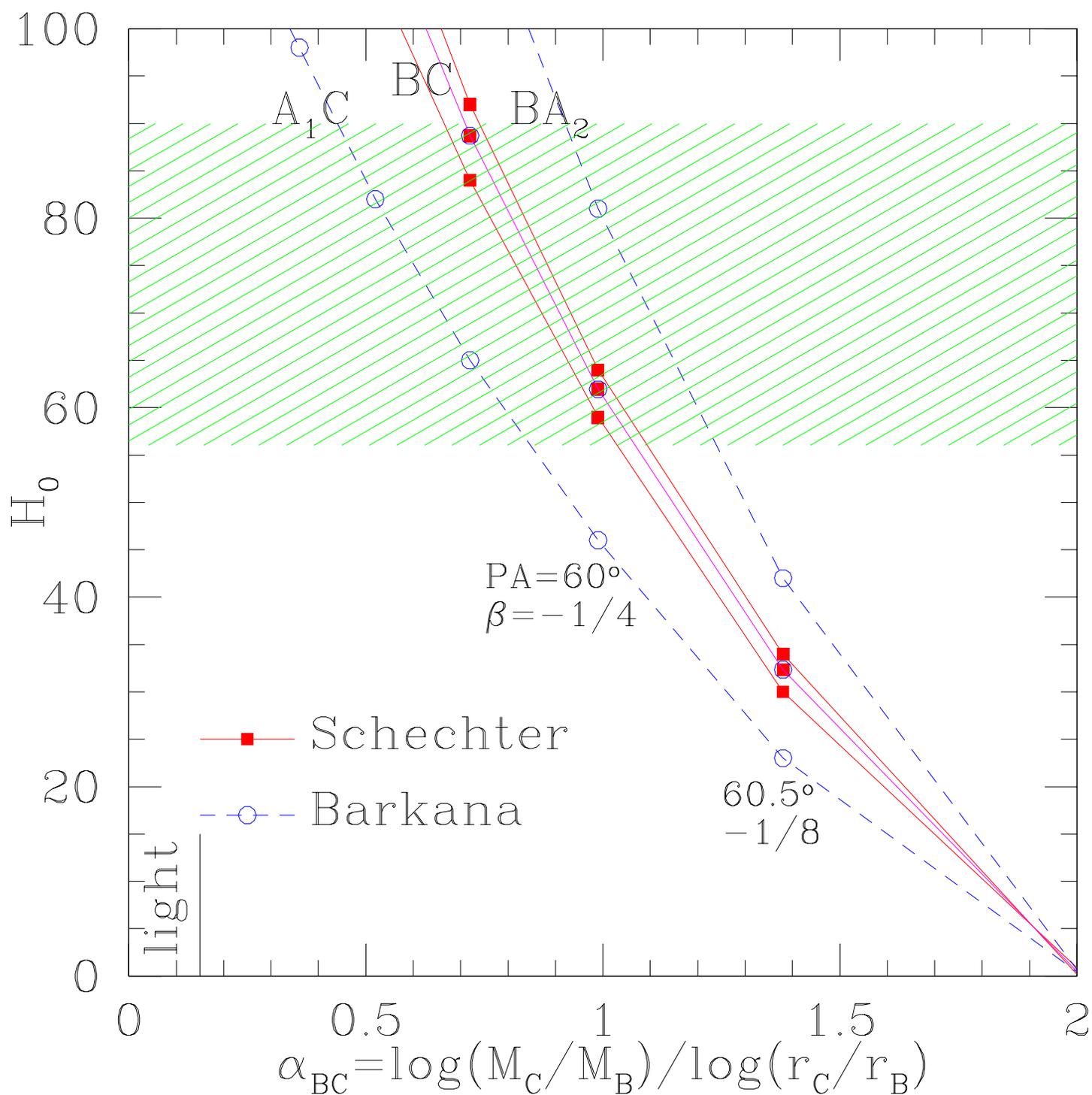}}
\caption{
The dependence of $H_0$ on the power-law slope $\alpha$,
where $\alpha$ is determined by the ratio of increase in mass
and increase in distance.
The short vertical line to the left shows the slope of increase
in light $\log (L_C/L_B)/\log (r_C/r_B)$.
The curves show the predictions for $H_0$ using $t_{AC}$,
$t_{BC}$, and $t_{BA}$ 
measured by Schechter et al. (1997), solid line, and Barkana (1997),
dashed line. The shaded region shows the consensus value for $H_0$.
}\label{hgraph}
\end{figure}

(vi) Our models do not yield a unique prediction of $H_0$, because it is
   sensitive to the lens power-law slope $\alpha$ (cf. Fig.~\ref{hgraph}),
   which appears to have a large spread, depending on how it is
   predicted.  The power-law slope is also sensitive to small
   uncertainties of the position angle of the lens $\theta_p$.  A
   change of $\theta_p$ by a few degrees from the observed value can
   change $\alpha$ from $0$ to $2$ and $H_0$ from 100 km/s/Mpc to 0.

(vii) Our models predict consistently a low time delay ratio $r_{ABC}$,
   which fits only the Schechter value.  This and the peculiar
   oscillation in the predicted density profiles, we believe, are due
   to the neglect of the external group.  This will be dealt with in
   detail in a follow-up paper. 

In summary, our semi-analytical lens models allow us to explore a
large range of physical lens density distributions.  We find that there
is a large systematic uncertainty in the lens models from fitting
image positions 
and time delays of PG1115+080 and isolated lens models are always
unsatisfactory for this quadruple system.

The authors thank Tim de Zeeuw for a careful reading of the manuscript
and the referee for a constructive report.
HSZ thanks Paul Schechter for discussions and encouragement.

\begin{table}
\caption{Effects of different cosmology on physical distance, velocity and 
time delay scales}\label{cosmotab}
\begin{tabular}{lrrrr}
($\Omega , \Lambda$) 	& (1.0 , 0 ) & (0.4 , 0.6 ) & (0.3 , 0.7) & (0.2 , 0.8)\\
kpc/$1''$ h   		& 2.808 & 3.127 & 3.195 & 3.271 \\
$V^2$ (km s$^{-1}$)$^2$/$1''$ 	& $316^2$ & $305^2$ & $303^2$ & $299^2$ \\
$\tau_{100}$ (days per sq. arcsec)  & 31.98 & 33.26 & 33.37 & 33.38 \\
\end{tabular}
\end{table}

\begin{table}
\caption{
Results for the models with the boxiness parameter
$\beta=-{1 \over 8}$ and
$\beta=-{1 \over 4}$}\label{betatab}
\begin{tabular}{lrrl}
Parameters & $\beta=-{1 \over 8}$ & $\beta=-{1 \over 4}$ & Comments \\
PA $\theta_p$ & $ 60.5^o$ & $ 60.0^o$ & Position angle of a principal axis of the potential (the group is at $\theta_g \sim 63^o$)\\
$(r_s,\theta_s)$ & $(0.09'',35^o)$ & $(0.03'',33^o)$ & Radius and position angle (counterclockwise from North) of the source \\
$q_\psi$ & $0.92$ & $0.86$ & Axis ratio of the potential  \\
$q_\kappa$	& 0.73-0.77 & 0.20-0.60 & Axis ratio of the surface density at typical radii of the images \\
$(\alpha_{BA},\alpha_{BC},\alpha_{AC})$	
&(1.1,1.4,1.6) & (0.51,0.99,1.31) & Effective power-law slope
$\alpha_{ij}={\log M_i-\log M_j \over \log \omega_i-\log \omega_j}$ from any two images $i$ and $j$\\
$M/L/h$		&27.0  & 26.3 & Mass-to-light ratio of the lens inside the image $C$ \\
$r_{ABC}$	&0.6 & 0.6 & Time delay ratio $t_{AC}/t_{BA}$ \\
$H_0$		&20-50 & 50-90 & Hubble constant in km/s/Mpc from Barkana (1997) time delay \\
\end{tabular}
\end{table}

{}


\begin{figure}
\epsfysize=10cm 
\centerline{\epsfbox{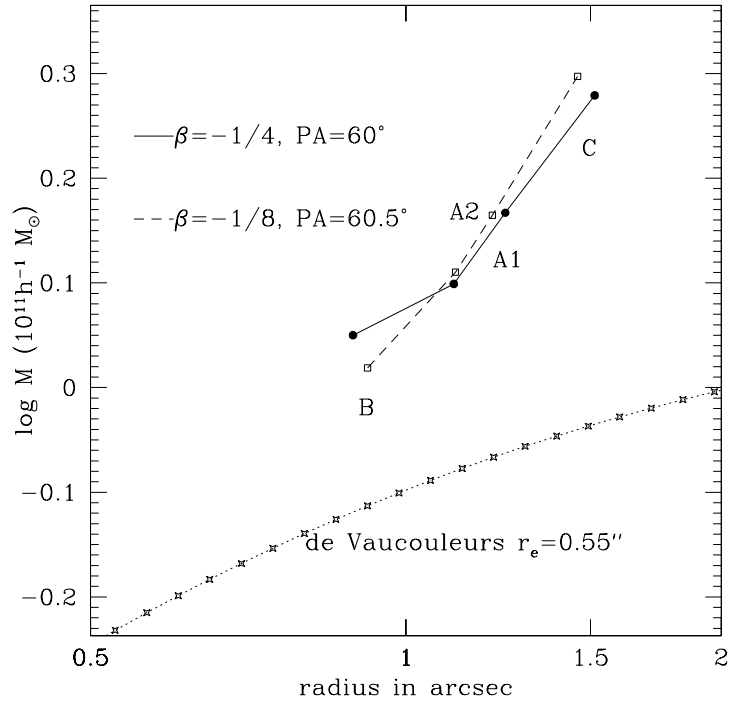}}
\caption{
The predicted lens mass (in units of $10^{11}h^{-1}M_\odot$)
inside the radii of the four images.
The predictions depend on the shape parameters of the lens, and are
made here for $\beta=-{1 \over 8}~ (\mbox{principal axis}
~\theta_p=60.5^o)$ and $\beta=-{1 \over 4}~ (\theta_p=60^o)$.
Also shown is the observed light profile (arbitrary flux normalization) with a 
de Vaucouleurs $r^{-{1 \over 4}}$-law and a half-light radius
of $0.55''$.  Note a radially-rising mass-to-light ratio.
}\label{massrad}
\end{figure}

\vfill \eject

\appendix
\section{Power-law models}

Within a certain range of radii, 
realistic lens mass distributions can be approximated by the following
bi-symmetric power-law of the radius $\omega$ 
\beq\label{powerlaw}
m(\omega) \equiv \omega {d\psi \over d\omega}
=b_0 r_0 \left({\omega \over r_0}\right)^\alpha,
~~~
\omega=r f(\theta)=r\left| 1-\delta \cos (2\theta-2\theta_p)\right|^\beta,
~~~r_0=1'',~~~0 < \alpha <2,
\eeq
where 
$b_0$ is the characteristic deflection strength at the characteristic
length scale $r_0$ (one arcsec) of the lens system, and $\theta_p$ defines
the principle axis of the lens. The corresponding lens potential is related to 
the enclosed mass $m(\omega)$ by
\beq
\psi(\omega) = {\rm const} + {m(\omega) \over \alpha}.
\eeq
The axis ratio (or its inverse) of the bi-symmetric potential
\beq\label{qpsi}
q_\psi \equiv \left({1-\delta \over 1+\delta}\right)^{\beta} \approx 1-\Delta,~~~\Delta=2\beta \delta,
\eeq
and the axis ratio of the density
\beq\label{qkappa}
q_\kappa \equiv 
\left({1-\delta \over 1+\delta}\right)^{1-\alpha \beta \over 2-\alpha}\left[{
1-({4\beta \over \alpha}-1)\delta \over 
1+({4\beta \over \alpha}-1)\delta }\right]^{1 \over 2-\alpha}
\approx 1-\Delta (1+2\alpha^{-1}),
\eeq
where the approximations are valid if the flattening is small.
The surface density $\kappa$ of the model is given by
\beq
\kappa = \frac{1}{2}b r^{\alpha-2} g^{\alpha\beta} ( \alpha g^2 + ( \beta^2\alpha - \beta) g'^2 + \beta g'' g),
\eeq
where $g\equiv 1-\delta\cos(2\theta-2\theta_p),
~~g'\equiv 2\delta\sin(2\theta-2\theta_p),
~~g''\equiv -4\delta^2\cos(2\theta-2\theta_p)$.

\section{Double-power-law models}

For the double power-law lens potential
\beq
\psi(\omega)=c_0 \left({\omega \over a_0}\right)^{\alpha_{in}}
\left[1+\left({\omega \over a_0}\right)^n\right]^{\alpha_{out}-\alpha_{in} \over n},
\eeq
the corresponding mass profile is given by
\beq
m(\omega)=\omega {d\psi(\omega) \over d \omega}
=\alpha(\omega) \psi(\omega),
\eeq
where
\beq
\alpha(\omega) = {d\log \psi(\omega) \over d \log \omega}
={\alpha_{in} + \alpha_{out}\left({\omega \over a_0}\right)^n \over
1+\left({\omega \over a_0}\right)^n }.
\eeq

Generally speaking, a double-power-law fit (cf. eq.~\ref{double}) 
involves searching for solutions
of the following equations in a five-parameter space 
$(n,\xi_0,\alpha_{out},\gamma_0,a_0)$:
\beq\label{doublefit}
\log m_i= \left[ \xi_0 + (1+\gamma_0)\alpha_{out} \log \omega_i \right]
-\left(1 + {\gamma_0 \alpha_{out} \over n} \right)
\log \left[1+\left({\omega \over a_0}\right)^n \right]
+\log\left[1+\gamma_0 + \left({\omega \over a_0}\right)^n\right],~~~i=1,2,3,4,
\eeq
where 
\beq
\gamma_0 \equiv { \alpha_{in} \over \alpha_{out} } -1,~~~
\xi_0 \equiv \log c_0  + \log \alpha_{out} -(1+\gamma_0)\alpha_{out} \log a_0,
\eeq
and $(\omega_i, m_i)$ for $i=1,2,3,4$ are the four data points 
in the mass-radius diagram (cf. Fig.~\ref{massrad}).
Note that the equations~\ref{doublefit} are linear in $(\xi_0,\alpha_{out})$, 
so these two variables can be eliminated by Gaussian substitution.
If we fix $n$, say $n=20$, then we are left with two equations
for two variables $(\gamma_0,a_0)$, which can be solved with a
two-dimensional iterative root finding routine. The solution converges
rapidly from an initial guess of $\gamma_0=0$ and $a_0=1''$.

\bsp
\label{lastpage}
\end{document}